\documentclass[fleqn,twoside]{article}
\usepackage{espcrc2}

\usepackage{epsfig}

\newcommand{\showpsfig}[2]{\centerline{\epsfig{file=#1,width=#2,clip=}}}
\title{Quark Loop Effects in Semileptonic Form Factors for Heavy-Light Mesons}
\author{ MILC Collaboration: C.~Bernard
\address{Department of Physics, Washington University, St.~Louis, MO 63130, USA},
C.~DeTar
\address{Physics Department, University of Utah, Salt Lake City, UT
  84112, USA}\thanks{Presented by C.~DeTar.},
Steven~Gottlieb
\address{Department of Physics, Indiana University, Bloomington, IN 47405, USA},
E.B.~Gregory
\address{Department of Physics, University of Arizona, Tucson, AZ 85721, USA}, 
U.M.~Heller
\address{American Physical Society, One Research Road, Box 9000, Ridge, NY 11961, USA},
C.~McNeile
\address{Department of Math Sciences, University of Liverpool, L69 3BX, UK},
J.~Osborn$\,\null^{\rm b}$,
R.L.~Sugar
\address{Department of Physics, University of California, Santa Barbara, CA 93106, USA},
and D.~Toussaint$\,\null^{\rm d}$
} 

\begin{document}
\begin{abstract}
We present preliminary results of a determination of the semileptonic
form factor for the decay of pseudoscalar heavy-light mesons to
pseudoscalar light-light mesons in full QCD.  In this preliminary
study we focus on the effects of dynamical quark loops.  Accordingly,
we compare results of simulations with matched quenched and Asqtad
dynamical gauge configurations.  The latter include three flavors of
light quarks.  Our simulation uses clover Wilson valence quarks,
treated in the Fermilab formalism.  Preliminary results, as yet
uncorrected by continuum matching factors, suggest a measurable
enhancement in the form factor due to dynamical quark loops over the
accessible range of $q^2$.
\end{abstract}

\maketitle
\section{INTRODUCTION}

Recent improvements in lattice determinations of $f_K$ and $f_\pi$
\cite{ref:precision} encourage us to hope that lattice values for
semileptonic form factors for heavy-light decays will achieve an
accuracy of a few percent in the near future.  Previous form factor
calculations have been done in the quenched approximation (except for
a preliminary study \cite{ref:lat99}), using light quarks with masses
of the order of the strange quark
\cite{ref:JLQCD,ref:NRQCD,ref:FNAL,ref:ROME}.  Here we report a
preliminary investigation of quenching effects by comparing results
from matched quenched and dynamical quark-loop gauge ensembles
\cite{ref:dataset,ref:Okamoto}.  The 409 $20^3\times 64$ quenched
lattices were generated with a one-loop Symanzik-improved gauge
action, and the 489 $20^3\times 64$ dynamical lattices were generated
with 2+1 flavors of Asqtad staggered quarks with bare quark masses
$am_{u,d} = 0.02$ and $am_s = 0.05$, the latter, approximately 1.2
times the physical strange quark mass.  The lattice scale $a$ was
based on the $\Upsilon$ 1S-1P mass splitting \cite{ref:upsilon}, giving
$a = 0.125$ fm for the quenched lattices and $a = 0.122$ fm for the
dynamical lattices.

Form factors were measured using clover-improved, propagating Wilson
light and heavy quarks.  Three light and five heavy quark masses were
used for each ensemble.  The mass of the lightest quark was
approximately equal to the strange quark mass, thus avoiding problems
with exceptional configurations.  Quark masses were tuned only
approximately between the two ensembles, so a comparison of results
requires some interpolation.  We included three heavy-light ``$B$''
meson three-momenta and 21 three-momentum transfers to the recoiling
light-light ``$\pi$'' meson.

\section{METHOD}

We measure the three point function in the usual way with a
light-light (or heavy-light) ``$\pi$'' interpolating operator ${\cal
O}_{\pi,G}$ at $t = 0$, a heavy-light ``$B$'' interpolating operator
${\cal O}_{B,G}$ at $t_f = 16$ and a weak vector current operator acting
over the range $t \in [0,16]$: 
\begin{displaymath}
  F_\mu({\bf k}{\bf q}{\bf p},t) =  
  \langle 0|O_{\pi, G}({\bf k},0) V_\mu({\bf q},t) O^\dagger_{B,E}({\bf p},t_f)|0\rangle
\end{displaymath}
The quark-antiquark wavefunction $G$ for the $\pi$ meson was a product
of two independent Gaussians of width $1.4a$ centered at the origin.
The quark-antiquark wavefunction $E$ for the $B$ meson was a plane
wave of c.m.~momentum ${\bf p}$ and a decaying exponential in the
relative coordinate.  The current vertex was projected onto
three-momentum ${\bf q}$, thus selecting a recoil three-momentum ${\bf
k} = {\bf p} - {\bf q}$.  Heavy-light and light-light two-point
functions were also measured over the same range of three-momenta
${\bf k}$ and with source/sink combinations $G$ and $E$:
\begin{eqnarray*}
  C_{X,GG}({\bf k},t) &=& 
  \langle 0|O_{X,G}({\bf k},0) O^\dagger_{X,G}({\bf k},t)|0\rangle \\
  &\rightarrow& Z_{X,G}({\bf k})Z_{X,G}({\bf k})e^{-E_\pi({\bf k})t} \\
  C_{B,GE}({\bf p},t) &=&  
  \langle 0|O_{B,G}({\bf p},0) O^\dagger_{B,E}({\bf p},t)|0\rangle \\
  &\rightarrow& Z_{B,G}({\bf p})Z_{B,E}({\bf p})e^{-E_B({\bf p})t}
\end{eqnarray*}
for $X = \pi,B$.  The form factor was extracted using the conventional
ratio method
\begin{displaymath}
 R_\mu(t) = \frac{F_\mu({\bf k},{\bf q},{\bf p},t)}
	{C_{\pi,GG}({\bf k},t)C_{B,GE}({\bf p},t_f-t)}
\end{displaymath}
\begin{displaymath}
\langle \pi({\bf k}) | V_\mu({\bf q}) | B({\bf p})\rangle = N Z_{\pi,G}({\bf k})Z_{B,G}({\bf p}) R_\mu,
\end{displaymath}
where $Z_{\pi,G}({\bf k})$ and $Z_{B,G}({\bf p})$ are hadron overlap
coefficients, $R_\mu$ is the plateau value of $R_\mu(t)$ and
\begin{displaymath}
  N  = N_{u,d}N_b = \sqrt{1 - \frac{3 \kappa_{u,d}}{4 \kappa_{\rm cr}}}
      \sqrt{1 - \frac{3 \kappa_b}{4 \kappa_{\rm cr}}}
\end{displaymath}
is the Fermilab normalization \cite{ref:EKM}.  Additional vertex
operators were measured to implement the tadpole improved rotation for
the quark wave functions in the vector current:
\begin{displaymath}
  q_I(x) = N_q \left[1 + a d \vec\gamma \cdot \vec D\right]q(x).
\end{displaymath}

For each quark mass combination the two-point functions
were fit simultaneously for all momenta to a ground plus first excited
state expression parameterized by a smooth function of momentum.  The
resulting ground state energies were then fixed when fitting the
three-point functions, and the resulting hadronic overlap coefficients
were used to complete the three-point function.

\begin{figure}[h]
\vspace*{-20mm}
\showpsfig{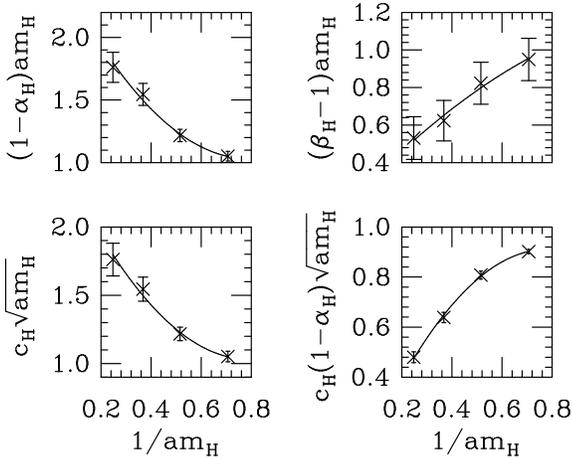}{100mm}
\vspace*{-18mm}
\caption{\label{fig:BKscaling}
  Quadratic heavy-quark scaling fits to the BK parameters for the
  quenched form factors with a strange spectator and recoil quark.}
\vspace*{-5mm}
\end{figure}

\begin{figure}[h]
\vspace*{-20mm}
\showpsfig{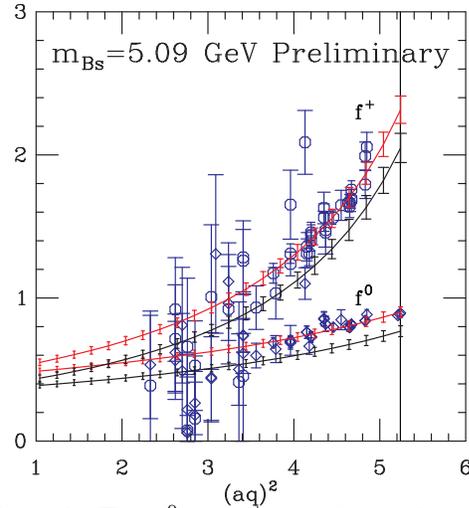}{100mm}
\vspace*{-18mm}
\caption{\label{fig: b_to_pi_ff2_zk0_sq2_bkfit} The $f^0$ and $f^+$
form factors in the presence of dynamical quark loops.  Here the
spectator quark and recoil quark have a mass equal to the strange
quark.  The mass of the heavy-light meson is approximately 5.09 GeV.
For each form factor the upper curve with error bars is a fit to the
simulation points and the lower curve is predicted from the quenched
ensemble. The prediction falls short by about 10\%.  }
\vspace*{-5mm}
\end{figure}

The Lorentz invariant form factors $f^+(q^2)$ and $f^-(q^2)$ were
extracted using kinematic factors based on physical hadron masses
determined from the two-point dispersion relations.  To compare
quenched and dynamical results, we required an interpolation in only
the heavy quark mass, since a light quark mass in each ensemble was
already matched to the strange quark mass, as determined from the
unmixed $s\bar s$ pseudoscalar meson.  To this end the form factors
were fit to a parameterization of Be\'cirevi\'c and Kaidalov (BK)
\cite{ref:BK}:
\begin{eqnarray*}
  f^+ &=& \frac{c_H(1 - \alpha_H)}
      {(1 - \tilde q^2)(1 - \alpha_H \tilde q^2)} \\
  f^0 &=& \frac{c_H(1 - \alpha_H)}
      {(1 - \tilde q^2/\beta_H)}
\end{eqnarray*}
($ \tilde q^2 = q^2/m_{B^*}^2 $), which incorporates constraints from
heavy quark scaling, the $B^*$ meson pole, and QCD sum rules.  The
resulting BK coefficients were then interpolated to the various
dynamical-ensemble physical heavy-light meson masses through a fit to
scaling forms as a function of the heavy-light meson mass, as shown in
Fig.~\ref{fig:BKscaling}.
A sample result is shown in Fig.~\ref{fig: b_to_pi_ff2_zk0_sq2_bkfit}.
The quenched prediction is about 10\% low.  At this heavy-light mass
the effect is greater than $2\sigma$.  We find a similar underestimate
for the entire range of heavy-light meson masses in this study,
from approximately $m_{D_s}$ to $m_{B_s}$ with light spectator
and recoil quarks at the strange quark mass.

\section{CONCLUSION} 

We have measured the semileptonic form factor for decays of a
heavy-light pseudoscalar meson to a light-light pseudoscalar meson
using improved Wilson valence quarks.  We chose carefully matched
quenched and Asqtad $2+1$ flavor staggered fermion ensembles to look
for the effects of virtual quarks.  Fixing all light quarks at the
strange quark mass, we required only a single heavy-quark
interpolation to make the comparison. For heavy-light masses of the
order of the $B_s$ mass, we found that virtual quark loops tend to
increase the form factors by approximately 10\%, a $>2\sigma$ effect.
Note however that our preliminary analysis neglects lattice/continuum
matching (lattice renormalization) of the heavy-light current. Since
the coupling constant is larger on the dynamical lattices, including
the renormalization may compensate for at least some of the difference
found between quenched and dynamical simulations.

This work is supported by the US NSF and DOE. Computations were
done with NPACI resources at SDSC and TACC and at PSC.  Gauge configurations
were generated at SDSC, ORNL, PSC, NCSA.



\begin{thebibliography}{}
  \bibitem{ref:precision}
C.~T.~Davies {\it et al.}  [HPQCD, MILC, UKQCD Collaborations],
arXiv:hep-lat/0304004.
  \bibitem{ref:lat99}
C.~W.~Bernard {\it et al.},
Nucl.\ Phys.\ Proc.\ Suppl.\  {\bf 83} (2000) 274.
  \bibitem{ref:JLQCD}
S.~Aoki {\it et al.}  [JLQCD Collaboration],
Phys.\ Rev.\ D {\bf 64} (2001) 114505.
  \bibitem{ref:NRQCD} 
J.~Shigemitsu, S.~Collins, C.~T.~Davies, J.~Hein, R.~R.~Horgan and G.~P.~Lepage,
Phys.\ Rev.\ D {\bf 66} (2002) 074506.
  \bibitem{ref:FNAL}
S.~M.~Ryan, A.~X.~El-Khadra, A.~S.~Kronfeld, P.~B.~Mackenzie and J.~N.~Simone,
Nucl.\ Phys.\ Proc.\ Suppl.\  {\bf 83} (2000) 328.
  \bibitem{ref:ROME}
A.~Abada, D.~Be\'cirevi\'c, P.~Boucaud, J.~P.~Leroy, V.~Lubicz and F.~Mescia,
Nucl.\ Phys.\ B {\bf 619} (2001) 565.
 \bibitem{ref:dataset}
C.~W.~Bernard {\it et al.},
Phys.\ Rev.\ D {\bf 64} (2001) 054506.
 \bibitem{ref:Okamoto} M.~Okamoto has carried out a companion
study of the same gauge ensembles with light staggered quarks, this
conference (2003).
 \bibitem{ref:upsilon} G.~P.~Lepage, private communication (2003).
  \bibitem{ref:EKM}
A.~X.~El-Khadra, A.~S.~Kronfeld and P.~B.~Mackenzie,
Phys.\ Rev.\ D {\bf 55} (1997) 3933.
 \bibitem{ref:BK}
D.~Be\'cirevi\'c and A.~B.~Kaidalov,
Phys.\ Lett.\ B {\bf 478} (2000) 417.
\end{thebibliography}
\end{document}